\documentclass[journal]{IEEEtran}
\usepackage{color}
\usepackage{import}
\usepackage{amsmath}
\usepackage{cite}
\usepackage{algorithmic}
\usepackage[ruled,vlined]{algorithm2e}
\usepackage{slashbox}
\usepackage{bm} 
\usepackage[pdftex]{graphicx}
\usepackage{amssymb}
\usepackage{multirow}
\usepackage{{physics}}
\usepackage[hidelinks]{hyperref}

\usepackage{xcolor}             

\newcommand{\lucas}[1]{\textcolor{blue}{#1}}

\usepackage{amsthm}  

\theoremstyle{definition}

\newcounter{cons}

\begin{document}
\title{Identification of NARX Models for \\ Compensation Design}

\author{Lucas~A.~Tavares, Petrus~E.~O.~G.~B.~Abreu and~Luis~A.~Aguirre
\thanks{L. A. Tavares and P. E. O. G. B. Abreu are with the Graduate Program in
	Electrical Engineering, Universidade Federal de Minas Gerais, Belo Horizonte, MG, Brazil
	(e-mail: amarallucas@ufmg.br; petrusabreu@ufmg.br).}
\thanks{L. A. Aguirre is with the Department
	of Electronic Engineering, Universidade Federal de Minas Gerais, Belo Horizonte, MG, Brazil
	(e-mail: aguirre@ufmg.br).}}


\maketitle


%

%
\IEEEpeerreviewmaketitle

\begin{abstract}
This report presents the modeling results for three systems, two numerical and one experimental.
In the numerical examples, we use mathematical models previously obtained in the literature as
the systems to be identified. The first numerical example is a heating system with a polynomial
nonlinearity that is described by a Hammerstein model. The second is a Bouc-Wen model that
represents the hysteretic behavior in a piezoelectric actuator. Finally, the experimental
example is a pneumatic valve that presents a variety of nonlinearities, including hysteresis.
For each example, a Nonlinear AutoRegressive model with eXogenous inputs (NARX) is identified
using two well-established techniques together, the  Error Reduction Ratio (ERR) method to hierarchically
select the regressors and the  Akaike's Information Criterion (AIC) to truncate the number of terms.
Using both approaches, the structure selection is achieved. The design of the excitation input is based
on preserving the frequencies of interest and force the system to achieve different points of operation.
Hence, having the structure previously selected with ERR and AIC, we use the Extended Least Squares (ELS)
algorithm to estimate the parameters. The results show that it is possible to identify the referred systems
with no more than five terms. These identified models will be used for nonlinearity compensation in future works.
\end{abstract}

\section{Introduction}

\IEEEPARstart{F}{or} many control applications, the first step is the identification of a representative
model to describe the fundamental aspects of the nonlinear dynamical system under investigation.
The identification process is usually addressed from a \textit{black-box and gray-box perspectives}.
In the black-box approach, the system is modeled exclusively from data without any prior knowledge
of the system. On the other hand, gray-box techniques use the data set and other information about
dynamical and static behaviors.

In this sense, it is necessary to define the mathematical representation of the models, such as
differential equations \cite{Lin_etal2013, liu2011intrachannel}, Radial Basis Functions (RBFs)
\cite{cao2018temperature,zhou2020dynamic,li2018position}, Neural Networks (NNs)
\cite{zhang2010neural,guo2018composite,meng2020neural}, and Nonlinear AutoRegressive models with
eXogenous inputs (NARX) \cite{Abreu_etal2020,Lacerda_etal2019}, among others. Having the representation
selected, the first step is to prepare the data set for identification purposes. In this sense,
it is essential to select an adequate sampling time, to excite the system at the frequencies of
interest and to achieve different many operation points \cite{Aguirre_2019}.

Being the data set carefully prepared and an appropriate mathematical framework defined,
it is important to select the model structure \cite{billings2013nonlinear}. The structure selection stage
defines which terms from a larger set of candidates will be part of the model. In the sequel, once the
structure is known, the model parameters are estimated with some optimization algorithm. The last step
is the validation of the model, which includes a different data set than the one used for identification \cite{Aguirre_2019}.

In this report, we present the modeling results for three systems, being the first two considered as numerical examples,
while the last is an experimental. In the numerical examples, we use mathematical models previously obtained in the literature as
the systems to be identified. The first numerical example is a heating system with a polynomial nonlinearity that is described by
a Hammerstein model. The second is a Bouc-Wen model that describes the hysteretic behavior in a piezoelectric actuator. Finally,
the experimental example is a pneumatic valve that presents a variety of nonlinearities, including hysteresis. For each example,
a NARX model is identified. The selection of an appropriate structure for the model is done considering the regressors and
the number of terms in the model \cite{billings2013nonlinear}. For this purpose, two well-established techniques are used together,
the Error Reduction Ratio (ERR) \cite{korenberg1988} method and the  Akaike's Information Criterion (AIC) \cite{Akaike1974}.
The ERR aims to sort the most representative regressors in the model, while the AIC truncates the number of terms. A relevant
part of this identification process is the design of the excitation input. It is based on preserving the frequencies of interest
and force the system to achieve different points of operation. Hence, having the structure previously selected with AIC and ERR,
we use the Extended Least Squares (ELS) \cite{ljung1987theory} algorithm to estimate the parameters. The results show that it is
possible to identify the referred systems with no more than five terms. In \cite{Tavares_2020_arxiv_comp}, these models will be used for compensation purposes.

\section{Background}
\label{back}

Being a nonlinear dynamical system $\cal S$ with a single input $u(t) {\in} \mathbb{R}$,
which determines a single output $y_s(t) {\in} \mathbb{R}$. Consider that a data set with
$N {\in} \mathbb{N}^+$ samples,  $Z^N {=} \{{u}(k), {y}_s(k)\}^{N}_{k=1}$, was collected from
$\cal S$ with a sampling time $T_{\rm s}$. The aim is to identify a Nonlinear Autoregressive
model with eXogenous inputs (NARX), namely $\cal M$, for representing the most relevant nonlinear
aspects of $\cal S$ \cite{leo_bil/85a}: 
\begin{eqnarray}
	\label{eq_model}
	y(k) {=}  f^\ell\big(y(k{-}1), \ldots ,y(k{-}n_y),u(k{-}\tau_{\rm d}), \ldots ,u(k{-}n_u)\big),
\end{eqnarray}

\noindent
where $y(k) \in \mathbb{R}$ is the deterministic model output that predicts $y_s(k)$, $f^\ell(\cdot)$
is a nonlinear polynomial function with degree $\ell \in \mathbb{N}^+$. $n_u,\,n_y \in \mathbb{N}^+$
are the maximum lags for $u$ and $y$, respectively, $\tau_{\rm d} \in \mathbb{N}^+$ is the pure time
delay. The $i$-th term $\psi_i(k-1)$ of $\cal M$ allows combinations up to degree $\ell$ of the regressor
variables $y(k-k_y)$, for $k_y \in \{1,\ldots,n_y\}$ and $u(k-k_u)$, for $k_u \in \{\tau_d,\ldots,n_u\}$:
\begin{equation}
	\label{ith_term}
	\psi_i(k-1)= \prod_{k_y=1}^{n_y}y(k-k_y)^{\ell_{{y},k_y}}\prod_{k_u=\tau_d}^{n_y}u(k-k_u)^{\ell_{u,k_u}},
\end{equation}

\noindent
where
\begin{equation}
\sum_{k_y=1}^{n_y}\ell_{y,k_y}+\sum_{k_u=\tau_d}^{n_u}\ell_{u,k_u} =\ell,	\nonumber
\end{equation}

\noindent
with $\ell_{y,k_y} \in \mathbb{N}$ and $\ell_{u,k_u} \in \mathbb{N}$. Model $\cal M$ is composed by
$n_{\theta} \in \mathbb{N}^+$ terms like (\ref{ith_term}), each one is multiplied by a constant parameter
$\hat{\theta}_i$. Thus, $\cal M$ can be expressed as:
\begin{equation}\label{eq_m_2}
	y(k) = \sum\limits_{i=1}^{n_\theta} \hat{\theta}_i\psi_i(k-1).
\end{equation}

Considering the modeling error of $\cal M$ or residues, $\xi(k)= y_{\rm s}(k)-y(k)$, it is possible
to rewrite (\ref{eq_m_2}) in a convenient form as follows:
\begin{equation}\label{vec_form}
	y_s(k) = \bm{\psi}^T(k-1)\bm{\hat{\theta}}+\xi(k),
\end{equation}
\noindent where
$\bm{\hat{\theta}}$ is the vector of parameters $[\hat{\theta}_1, \ldots, \hat{\theta}_{\rm n_{\theta}}]^T$,  
$\bm{\psi}$\hbox{$(k-1)$} is the set of regressors (\ref{ith_term}) organized in a vector $[\psi_1$\hbox{$(k-1)$},
$\ldots, \psi_{n_{\rm \theta}}(k-1)]^T$ and $T$ indicates the transpose. Applying (\ref{vec_form}), for $k=1,\ldots,N$, over the data set $Z^N$, it is possible to build the following
equation in matrix form \cite{Aguirre_2019}:
\begin{equation}\label{psi}
\bm{y}_s = \bm{\Psi}\bm{\hat{\theta}}+\bm{\xi},
\end{equation}

\noindent
where $\bm{\xi} = [\xi(1), \ldots, \xi(N)]^T$, $\bm{y}_s = [y_s(1), \ldots, y_s(N)]^T$,
$\bm{\psi_i} = [\psi_i(1), \ldots, \psi_i(N)]^T$ for $i\in \{1,\ldots,n_\theta\}$, and
$\bm{\Psi} \in \mathbb{R}^{N\times n_{\theta}}$ is the regressor matrix given by
$\bm{\Psi}=[\bm{\psi}_i(1), \ldots, \bm{\psi}_i(N)]^T$. As (\ref{psi}) is linear
in the parameters, classic regression methods can be used, e.g., Least Squares (LS).
The LS method provides the following parameter vector $\bm{\hat{\theta}}_{\rm LS}$ optimal,
in the sense of least squares of the modeling error:
\begin{equation}\label{eq_MQ}
	\bm{\hat{\theta}}_{\rm LS} = [\bm{\Psi}^{T}\bm{\Psi}]^{-1}\bm{\Psi}^{T}\bm{y}_s.
\end{equation}

When $\xi(k)$ is autocorrelated, the LS method presents a polarized response. This is
common in the presence of noise, for which the addition of {\it moving average} (MA)
terms is an alternative to solve such a problem. However, this addition provides a
nonlinear model in the parameters, and the classic LS cannot be used. A way to
circumvent this issue is given by the Extended Least Squares (ELS) \cite{billings2013nonlinear}.
This iterative algorithm uses $\xi(k)$ from the previous iteration to extend the regressor matrix,
being executed until a convergence criterion is reached, as illustrated by the following steps
\cite{ljung1987theory}:
\begin{enumerate}
	\item{Based on Eq. \ref{eq_MQ}, estimate $\bm{\hat{\theta}}_{LS}$;}
	\item{Calculate the vector of residues, $\bm{\xi}_1 = \bm{y}_s - \bm{\Psi\hat{\theta}}_{\rm LS}$;}
	\item{Define the iteration $i=2$ and the convergence limit $\zeta \in \mathbb{R}^+$;}
	\item{Build ${\bm{\tilde\Psi}_{i}} = [\bm{\Psi}$ $\scalebox{0.75}{\vdots}$$\text{  } \bm{\xi}_{i-1}]$, the extended regression matrix;}
	\item{Estimate via LS the new vector of parameters at each iteration: 
		$\bm{\hat{\theta}}_{{\rm ELS}_i} =  [{\bm{\tilde\Psi}_{i}}^{T}\bm{\tilde\Psi}_{i}]^{-1}{\bm{\tilde\Psi}_{i}}^{T}\bm{y}_s$};
	\item{Determine the current residues: 
		$\bm{\xi}_{i} = \bm{y}_s - \bm{\tilde\Psi_{i}\hat{\theta}}_{{\rm ELS}_i}$;}
	\item{If ${||\bm{\hat{\theta}}_{{\rm ELS}_i}-\bm{\hat{\theta}}_{{\rm ELS}_{i-1}}||}_2 < \zeta$, make $\bm{\hat{\theta}}_{\rm ELS} = \bm{\hat{\theta}}_{{\rm ELS}_i}$ and finish the process. Otherwise, make $i=i+1$ and return to step $4$. $||\bullet||_2$ is the quadratic norm.}
\end{enumerate}

The described technique to determine the parameters must be applied to a previously identified structure.
The selection of this structure for $\cal M$ can be done with the following techniques used together.

\subsection{The Error Reduction Ratio Method}

The Error Reduction Ratio (ERR) is an optimization-based technique that quantifies the contribution
of each term to explain the variance of the modeling error \cite{korenberg1988}. For this application,
each model regressor must be orthogonal to the data that can be achieved with Householder transformation
\cite{householder1958unitary}. Taking the average value of $y_s(k)^2$ on the data, it is possible to find
the following expression \cite{billings2013nonlinear}:
\begin{equation}
	\dfrac{1}{N}\sum\limits_{k=1}^{N}y_s(k)^2 = \sum\limits_{i=1}^{n_\theta}{\hat{\theta}_i}^2{\bm {{w_{i}}^{T}}}{\bm {w_i}} + {\bm {\xi^T}}{\bm {\xi}}.
\end{equation}

The {\it error reduction ratio} due to the inclusion of $i$-th regressor is given by the following expression:
\begin{equation}
	|ERR_i| = \dfrac{{\hat{\theta}_i}^2{\bm {{w_{i}}^{T}}}{\bm {{w_i}}}}{\bm {y^Ty}}.
\end{equation}

\noindent
The inclusion of each term in the model reduces the variance of the modeling error. The index $|ERR_i|$ quantifies
the contribution of the $i$-th term in the reduction of this variance. By selecting the regressors that provide the
highest values for this index, it is possible to reduce the candidate set. This is relevant because the number of regressor candidates tends to considerably increase with the
raise of $\ell$ \cite{billings2013nonlinear}. 
However, this technique only sorts the candidate terms hierarchically, it is still necessary to choose the number of candidate terms.

\subsection{The Akaike's Information Criterion}
\label{aic_cap2}

The Akaike's Information Criterion (AIC) is a tool that helps to define the number of terms to be included in the model,
based on minimizing a cost function \cite{Akaike1974}. The addition of new terms increases the model complexity, which
allows a better fit. Considering that the terms were sorted with ERR, a given number of terms $n_\theta$ is enough to determine
${\cal M}_{n_{\theta}}$ with an output $y_{n_\theta}(k)$. The polarized error is the variance of the residues of this model, $\sigma_{\xi}^2(n_\theta)=\sigma^2[y_s(k)-y_{n_\theta}(k)]$. When $n_\theta$ increases, $\sigma_{\xi}^2(n_\theta)$ tends to reduce,
although increases in complexity do not guarantee a better generalization performance. Thus, this criterion aims to penalize the
rise in the number of terms. The AIC cost function is:
\begin{equation}\label{aic_eq}
	J_{AIC(n_\theta)} = N{\rm ln}[\sigma_{erro}^2(n_{\theta})] + 2n_{\theta}.
\end{equation}

Hence, this function tends to achieve a minimum value for some value of $n_\theta$. It is convenient to highlight that this is a purely statistical criterion that helps in the identification process. Thus, there is no guarantee that the obtained model is valid in the identification perspective.

\subsection{NARX Hysteresis Modeling}
\label{sub_comp_hys}

Sufficient conditions for NARX models to mimic hysteresis loops were presented by \cite{Martins_Aguirre2016}.
The authors have shown that with the inclusion of the first difference of the input $\phi_{1}(k){=}u(k){-}u(k{-}1)$
and its corresponding sign function, $\phi_{2}(k) {=} {\rm sign}(\phi_{1}(k))$, it is possible to represent
a hysteretic behavior for loading-unloading inputs. A general extended deterministic NARX model set \cite{bil_che/89}
referred as ${\cal{M}}_{\rm h}$ is represented by:
\begin{eqnarray}
	\label{m_narx_hys}
	\!\!\!\!\!y(k) &{=}& g^{\ell}\big(y({k-1}),\cdots,y(k-n_{y}), \,u(k-\tau_{\rm d}),\cdots, \nonumber \\ 
	&{~}& \hspace{3mm} u(k-n_u), \phi_{1}(k-1), \,\phi_{2}(k-1) \big), 
\end{eqnarray}

\noindent
where $g^{\ell}(\cdot)$ is a polynomial function of the regressor variables up to degree $\ell$, and the other
parameters are previously mentioned. Model ${\cal M}_{\rm h}$ \eqref{m_narx_hys} presents two sets of equilibria
under loading-unloading inputs: one for loading with $\phi_{2}(k){=}1$, and one for unloading with $\phi_{2}(k){=}{-}1$
\cite{Martins_Aguirre2016, Abreu_etal2020}.

It has been shown that for ${\cal{M}}_{\rm h}$, some regressors can be excluded, regardless of the lags $\tau_u$ and
$\tau_y$ \cite{Abreu_etal2020}, as summarized below:
\begin{itemize}
	\item[(i)] $y^{p}(k-\tau_y)$,  $y^{p}(k-\tau_y)\phi_1(k-\tau_u)^q$, 
	\\$y^{p}(k-\tau_y)\phi_2(k-\tau_u)^q$
	for $p{>}1,~\forall q$, \cite{Aguirre_Mendes1996},
	\item[(ii)] $\phi_2^q(k-\tau_u)$ for $q>1$ \cite{Martins_Aguirre2016}, and
	\item[(iii)] $y^{p}(k-\tau_y)u^{m}(k-\tau_u)$ and $u^{m}(k-\tau_u)$, $\forall p, m$ \cite{Abreu_etal2020}. 
\end{itemize}

It is important to note that such constraints are specified for the identification of ${\cal{M}}_{\rm h}$,
as detailed in the references. Considering $\Sigma_y$, as the sum of parameters of all linear output regressors,
if a model is identified with these constraints and forcing the parameters to generate $\Sigma_y=1$, the model
remains in the last state when the reference becomes constant. This property is desired	for hysteresis models
and can be achieved using a constrained least squares estimator, as detailed in \cite{Abreu_etal2020}.

\section{The Identification Methodology}
\label{sp}

The procedure proposed here takes into account the need to excite the system in a wide frequency range and reach a
variety of operating points \cite{schoukens2019}. Therefore, the procedure starts with the need to define the
frequencies, $f_i \in \mathbb{R}^{+}$ for $i{=}1, \ldots, n$, that will be preserved in the signal. For instance,
in the case of hysteretic systems, it is necessary to preserve low-frequency information in the identification
data \cite{Ikhouane_Rodellar2007}. To circumvent this type of problem, the fifth-order low-pass Butterworth filters,
${\cal H}_{i}(q)$, are designed with a cutoff frequency that concentrates the spectral power of interest. In this work,
we consider that these filters are applied to a purely random signal $e_i(k){=}{\cal N}(0,\,1),~k=1,\,2, \ldots N_i$
where ${\cal N}(0,\,1)$ is a standard normal distribution. Each signal $e_i(k)$ is conditioned to have $1$ as maximum
and $-1$ as minimum, given by:
\begin{equation}
	\label{Eq:RandomSignal}
	e_i(k) = 2 \left (\dfrac{e_i(k)-{\rm min}[e_i(k)]}{{\rm max}[e_i(k)]-{\rm min}[e_i(k)]}\right)-1,
\end{equation}

\noindent
and its number of samples $N_i$ is specified to have an input signal $u(k)$ with $N$ samples, such that
$N{=}N_{1}{+}N_{2}{+}\cdots {+}N_{n}$. As will be seen, the input signal $u(k)$ is obtained by concatenating
a signal $s_i(k)$, for $i{=}1, \ldots, n$.

To ensure that the signal $u(k)$ can achieve a variety of operating regions, it is considered
the operations points as $o_j \in \mathbb{R}$ for $j{=}1, \ldots, v$. For each operation point $o_j$,
it is defined an amplitude $G_j$ around this particular point. In addition, the values assigned for
each number of samples $N_i$ must be defined as multiples of the number of operating points, $v$.
This is necessary since it is intended that each frequency has information around all operating points.
In this way, the signal $s_i(k)$ that is a part of $u(k)$ can be produced as follows:
\begin{align}
	\label{eq_s}
	s_i(k){=} \left\{
	\begin{array}{ll}
		\!\alpha_{i,1}{\cal H}_i(q)e_i(k) + o_1, &  k{=}1, \, 2, \ldots \dfrac{N_i}{v} \\ 
		\!\vdots &\vdots
		\\
		\!\alpha_{i,v}{\cal H}_i(q)e_i(k) + o_v, &   k=(v{-}1)\dfrac{N_i}{v}{+}1,\\ & \hspace{4mm}(v{-}1)\dfrac{N_i}{v}{+}2, \ldots, N_i. \\
	\end{array} \right.
\end{align}

The constants $\alpha_{i,j}$ are defined to limit the excursion of the input around each operating point.
For a given operating point $o_j$, we have $\alpha_{i,j}{=}G_j/{\rm{max}}[e_i(k_{i,j})]$ for which
$k_{i,j} = (j-1)(N_i/v)+1,(j-1)(N_i/v)+2, \ldots, j(N_i/v)$. Applying  \eqref{eq_s} for $i=1,2,\ldots,n$,
we get $s_1, s_2, \ldots, s_n$ and the $u(k)$ is the concatenation of these inputs:
\begin{equation}
	u(k) = [s_1 \quad s_2 \quad \ldots \quad s_n].
	\label{eq_u}
\end{equation}

\noindent
Finally, selecting the maximum frequency $f_i$, the low-pass filter related to this frequency is used to
eliminate possible higher frequencies originating from concatenation.

This procedure is performed twice, in order to have different realizations for identification and validation inputs,
which are applied to the system for data collection. To achieve more robust experiments, Gaussian noise was added
directly to the output to obtain $\sigma_n/\sigma_s = 5\%$, where $\sigma_n$ is the standard deviation of the noise, and $\sigma_s$ is the standard deviation of the signal. To complement our validation data, we also
generate sinusoidal inputs at different frequencies and amplitudes.

Considering the model structure selection step, the ERR and AIC tools are used together, as detailed in
Sec.\,\ref{back}. First, the candidate regressors to compose the NARX model are evaluated and ranked in
such way that the ERR criterion is minimized. Then, the AIC is used to truncate the number of regressors
in the final model. For the estimation of parameters, in this paper, the ELS method is used.

\section{Examples}
\label{NumericalExamples}

This section illustrates the identification procedure proposed in Sec.\,\ref{sp} for two simulated benchmark
systems and one experimental. The first one is a heating system modeled by a Hammerstein model with a polynomial
nonlinearity, and the second is a Bouc-Wen that aims to model the hysteresis of a piezoelectric (PZT) device.
The experimental example is a pneumatic valve. All examples are treated as systems to be modeled and these models
are used for compensation in \cite{Tavares_2020_arxiv_comp}. To evaluate the performance of the prediction and
compensation achieved, the mean absolute percentage error ({\rm MAPE}) index is computed as follows:
\begin{equation}
	\label{Eq:MAPE}
	{\rm{MAPE}} = \dfrac{\sum_{k=1}^{N}|y(k) - \hat{y}(k)|}{N|{\max}({\bm y})-{\min}({\bm y})|}.
\end{equation}

\subsection{A Heating System}
\label{sec_hs_sys}

Consider the bench test system, a small electrical heater, modeled by the following
Hammerstein model \cite{agu_eal/05iee}:
\begin{align}
	\label{heat_sys_s}
	y(k)  = & \beta_1 y(k-1) + \beta_2 v(k-1) + \beta_3 y(k-2) + \beta_4 v(k-2), \nonumber \\ 
	v(k)  = & p_1 u(k)^2+ p_2 u(k), 
\end{align}  

\noindent 
where $y(k)$ is the normalized temperature, and $u(k)$ is the electric power applied to the heater within the range
$ 0{ \leq} u(k) {\leq} 1$. The data set of such a system is the same considered in \cite{agu_eal/02iee}, which is
available at \lucas{\textcolor{blue}{\url{https://bit.ly/3iQ6rCF}}}.
Here, we reestimate the parameters of model (\ref{heat_sys_s}), obtaining: $p_1 = 4.639331 \times 10^{-1}$,
$p_2 = 5.435865 \times 10^{-2}$, $\beta_1=1.205445$, $\beta_2 = 8.985133 \times 10^{-2}$,
$\beta_3= -3.0877507 \times 10^{-1}$ and $\beta_4 = 9.462358  \times  10^{-3}$. This reestimation was need because the parameters provided 
by \cite{agu_eal/05iee} have few decimal digits, which severely impact in the results for discrete systems.

\begin{figure}[htb]
	\centering
	\includegraphics[width=1\columnwidth]{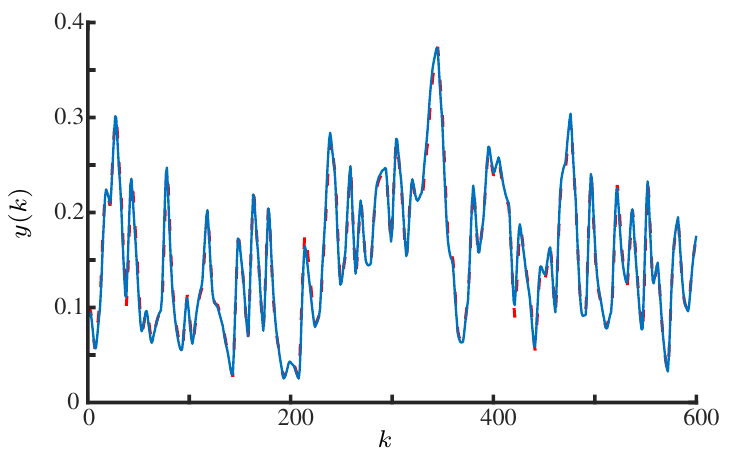}
	 \vspace{-0.7cm}
	\caption{Performance of \eqref{heat_sys_s} over measured validation data provided by \cite{agu_eal/02iee}. Continuous line (\textcolor{blue}{--}) is the measured data, while the dashed line (\textcolor{red}{- -}) is the free-run
	simulation of model \eqref{heat_sys_s}. }
	\label{fig_valid_hs_sys}
	\vspace{-0cm}
\end{figure}

The independent term in the second-order polynomial that gives $v(k)$ was omitted
in order to ensure $y=0$ when $u=0$.  $p_1$ and $p_2$ were estimated directly from 
static data using LS, although $\beta$'s were estimated from dynamical data using ELS
algorithm. The operation region of the model is $u(k) \in [0,~1]$ and $y(k) \in [0,~0.5]$.
The validation result (Fig.\,\ref{fig_valid_hs_sys}) shows the performance of the model given by \eqref{heat_sys_s}. 

\begin{figure}[htb]
	\centering
	 \includegraphics[width=1\columnwidth]{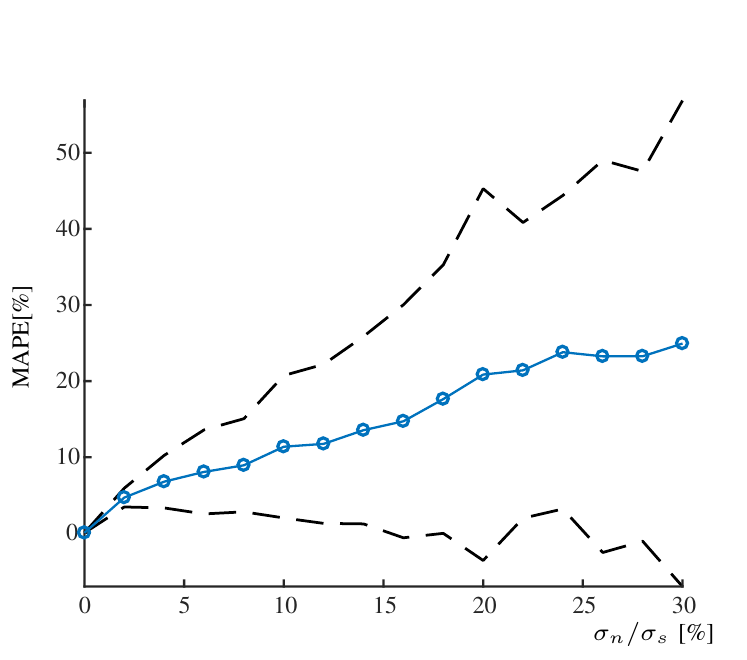}
	 \vspace{-0.75cm}
	\caption{Monte Carlo tests are performed to evaluate the MAPE [\%] variation in function of the raising from $\sigma_n/\sigma_s$ [\%] where $\sigma_n$ is the standard deviation of the noise, and $\sigma_s$ is the standard deviation of the signal. For each $\sigma_n/\sigma_s$ value, 50 Monte Carlo tests are performed. We show the mean of 50 tests (\textcolor{blue}{$\ominus$}) and $\pm$2 standard deviations (\textcolor{black}{- -}). The continues lines (\textcolor{blue}{--}) are only illustrative, since the tests were taken for successive raises of 2\% in $\sigma_n/\sigma_s$.}
	\label{fig_hs_ruido}
	\vspace{-0.1cm}
\end{figure}

From now on, the Hammerstein model (\ref{heat_sys_s}) will be treated as the system $\cal{S}$
to be identified with a NARX polynomial model $\cal{M}$ (\ref{eq_model}). The input $u(k)$ was
designed following (\ref{eq_s}) and (\ref{eq_u}) with: $n=2$, $f_1=0.001 {\rm{Hz}}$, $f_2=0.005 {\rm{Hz}}$,
$v=3$, $o_1=0.3 {\rm{V}}$, $o_2=0.5 {\rm{V}}$, $o_3=0.7 {\rm{V}}$, $G_1=G_2=G_3=0.2 {\rm V}$, $N=2000$, $N_1=1000$,
and $N_2 = 1000$. Figure\,\ref{fig_hs_ruido} shows the results obtained with Monte Carlo Tests to assess
how the predictive capacity of the identified models degrade with the increase the noise power, using the MAPE index.
Considering the system output $y_s(k)$, for each Monte Carlo test, a Gaussian noise is added with a given value of the ratio $\sigma_n/\sigma_s$, where $\sigma_n$ is the standard deviation of the noise, and $\sigma_s$ is the standard deviation of the signal.
The $\sigma_n/\sigma_s$ values vary from $0\%$ to $30\%$ with successive increase of $2\%$. For each $\sigma_n/\sigma_s$ value, $50$ perturbed output
signals are generated and their respective models are identified taking into account the previous recommendations.
Finally, we get the mean and standard deviation of MAPE for each $\sigma_n/\sigma_s$ value. From Fig.\,\ref{fig_hs_ruido} is possible
to see that the model quality of the identification is, in average, more frequently affected for raising
values of $\sigma_n/\sigma_s$. Also, the predictability is harmed, since there is an enlargement in the confidence intervals.

	 \vspace{-1.3cm}

\begin{figure}[!h]
	\centering
	\includegraphics[width=0.95\columnwidth]{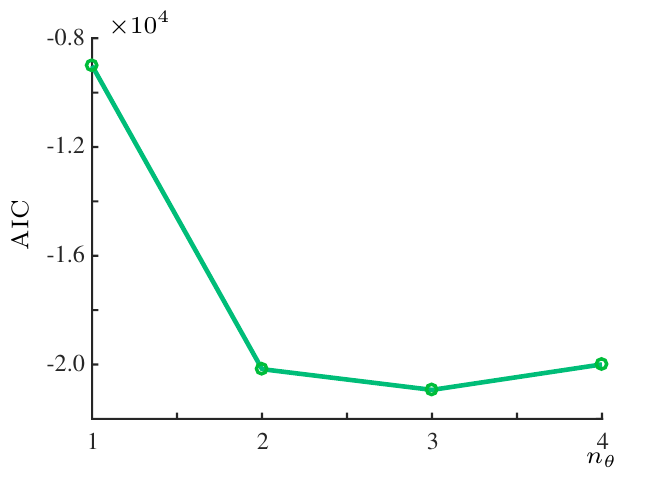}
	\vspace{-0.4cm}
	\caption{AIC criterion used for the definition of $n_{\theta}$. It has been chosen the model of 3 terms due to the index minimization. The continuos line have no meaning, it was used for facilitates the visualization.} 
	\label{fig_hs_aic}
	\vspace{-0.1cm}
\end{figure}

As aforementioned, in the examples used, Gaussian noise was added directly to the output to yield a $\sigma_n/\sigma_s=5\%$.
Taking $\ell=3$ and $n_y = n_u = 3$, the ERR method selects the more representative regressors and the AIC criterion
is calculated as shown in Fig.\,\ref{fig_hs_aic}.

\begin{figure}[htb]
	\centering
	\includegraphics[width=1\columnwidth]{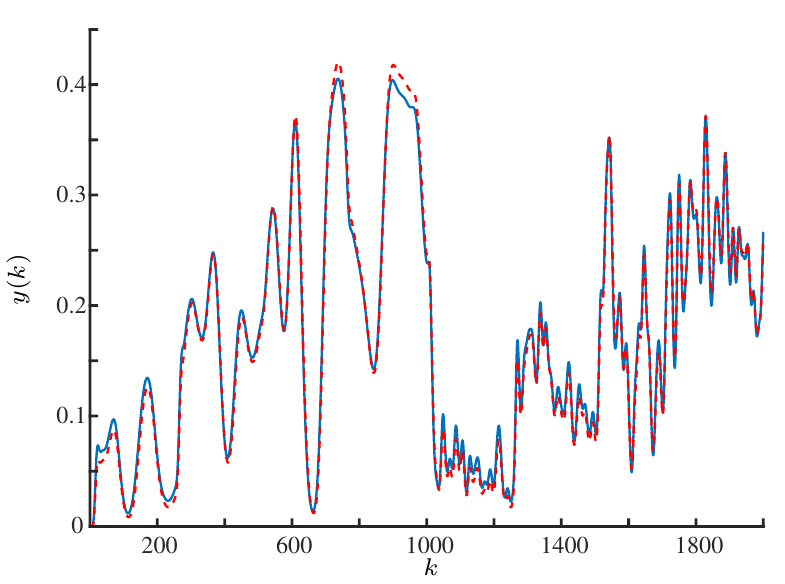}
	\vspace{-0.85cm}
	\vspace{0cm}
	\caption{Noise-free validation data for model $\cal M$ \eqref{hs_model}: continuous line (\textcolor{blue}{--}) presents values of $y(k)$ for the system $\cal S$ \eqref{heat_sys_s}, while the dashed line (\textcolor{red}{- -})  is the free-run
	simulation of model \eqref{hs_model}. }
	\label{fig_hs_val_m}
	\vspace{-0.2cm}
\end{figure}

By minimizing the AIC criterion, the following three-term model  $\cal M$ was obtained according to the
procedure detailed in Sec\,\ref{sp}:
\begin{equation}
\label{hs_model}
y(k) = \hat{\theta}_1y(k-1)+\hat{\theta}_2 u(k-2)^2 + \hat{\theta}_3y(k-2), 
\end{equation}

\noindent 
where $\hat{\theta}_1 = 8.958185 \times 10^{-1}$, $ \hat{\theta}_2 = 6.393347 \times 10^{-2}$, and
$\hat{\theta}_3 = -1.746750 \times 10^{-2}$. The free-noise results for the validation data set are
shown in Fig.\,\ref{fig_hs_val_m}. A more detailed assessment of this model for sinusoidal inputs can
be found in Table I in \cite{Tavares_2020_arxiv_comp}. The mentioned work provide an compensation approach
based on this identified model.

\subsection{A Hysteretic System}
\label{sec_bw}

In this example, the following Bouc-Wen model was used to describe the hysteretic behavior of a
piezoelectric actuator (PZT) that is an unimorph cantilever \cite{Rakotondrabe2011}:
\begin{align}
\dot{h}(t) =& \alpha_{\rm bw}\dot{u}(t)-\beta_{\rm bw}|\dot{u}(t)|h(t)-\gamma_{\rm bw} \dot{u}(t)|h(t)|,  \nonumber\\
y(t) =& \nu_{y}u(t) - h(t),
 \label{eq_bw}
\end{align}

\noindent
where $u(t)$[V] is the voltage input, $y(t)$[$\rm{\mu}m$] is the position output,
the parameters $\alpha_{\rm bw} = 0.9 [\mu\rm{ m}/\rm{V}]$ and
$\beta_{\rm bw}=\gamma_{\rm bw} = 0.008[\rm{V}^{-1}]$ determine the hysteresis loop, while
$\nu_{y}=1.6 [\mu\rm{ m}/\rm{V}]$ is a weight factor for the output. Here,
\eqref{eq_bw} is referred as the system $\cal S$ to be identified, which is simulated
with a fourth-order Runge-Kutta method considering the integration step $\delta_t=5 \rm{ms}$.

\begin{figure}[htb]
	\centering
	\includegraphics[width=1\columnwidth]{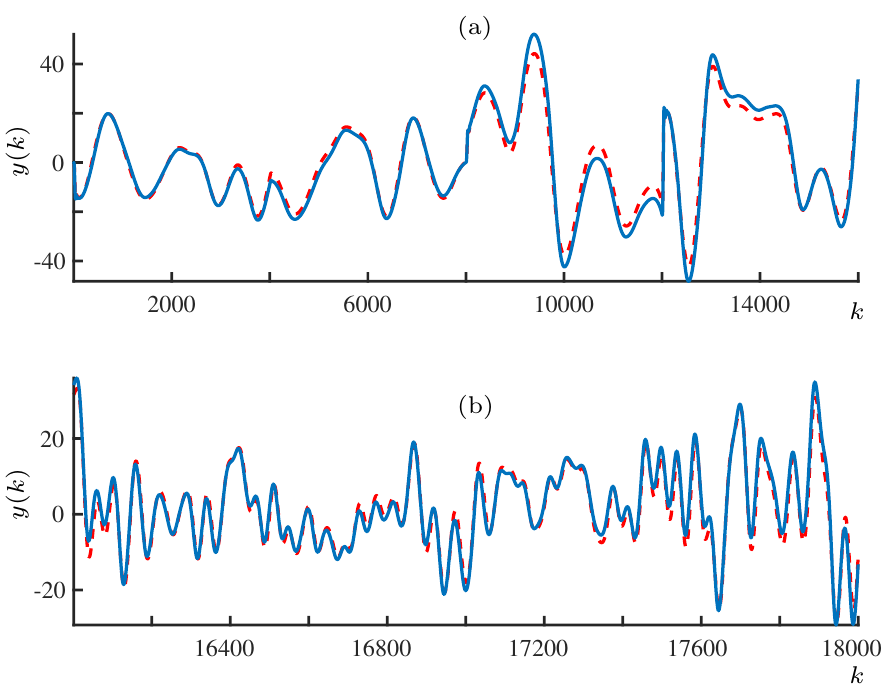}
	\vspace{-0.65cm}
	\caption{
		\hspace{-0.32cm}Noise-free validation data for model ${\cal M}_{\rm h}$ (\ref{bw_model}): continuous line (\textcolor{blue}{--}) presents values of $y(k)$ for the system $\cal S$ \eqref{eq_bw}; dashed line (\textcolor{red}{- -}) presents output values of the model \eqref{bw_model}. In (a), we have the temporal evolution of the outputs for lower frequency $f_1 =0.2 \rm{Hz}$ $(0 \leq k\leq 16000)$, while in (b) is shown the first 3000 samples for higher frenquency $f_2 =5 \rm{Hz}$, $(16001 \leq k\leq 18000)$.}
	\label{fig_bw_valid}
\end{figure}

To identify a NARX polynomial model ${\cal M}_{\rm h}$ (\ref{m_narx_hys}) for $\cal S$,
the identification input $u(k)$ is designed according to (\ref{eq_s}) and (\ref{eq_u})
with: $n=2$, $f_1=0.2 {\rm{Hz}}$, $f_2=5 {\rm{Hz}}$, $v=2$, $o_1=o_2= 0 {\rm{V}}$,
$G_1 = 25 {\rm V}$,  $G_2 =50 {\rm V}$, $N = 19200$, $N_1 = 16000$ and $N_2 = 3200$. 
Gaussian noise was added directly to the output to yield a $\sigma_n/\sigma_s=5\%$, as in the previous example. 
At higher noise levels the identified models are somewhat worse, as expected, similarly the
discussion previous done in Sec.\ref{sec_hs_sys}. Based on Sec.\,\ref{sub_comp_hys}, $u(k)$,
$y(k)$, $\phi_{1}(k){=}u(k){-}u(k{-}1)$, and $\phi_{2}(k) {=} {\rm sign}(\phi_{1}(k))$ are
chosen as candidate regressors, taking $\ell = 3$ and $n_y = n_u = 1$ the following
forth-term model  ${\cal M}_{\rm h}$ was obtained from AIC and ERR to select the structure,
and ELS for parameter estimation:
\begin{eqnarray}
\label{bw_model}
y(k)\!\!\!\! &{=}&\!\!\!\!\hat{\theta}_1y(k{-}1){+}\hat{\theta}_2\phi_{2}(k{-}1)\phi_{3}(k{-}1)u(k{-}1) \nonumber \\
             & & {+}\hat{\theta}_3\phi_{2}(k{-}1)\phi_{3}(k{-}1)y(k{-}1){+}\hat{\theta}_4\phi_{2}(k{-}1), 
\end{eqnarray}

\noindent
where $\hat{\theta}_1 = 1.000099$, $ \hat{\theta}_2 = 6.630567 \times 10^{-3}$, $\hat{\theta}_3 = -6.247018 \times 10^{-3}$,
and $\hat{\theta}_4=7.892915$. Note that this model has a structure which consent with the restrictions (i), (ii), and (iii)
of Sec.\,\ref{sub_comp_hys}. However, this model does not comply with the constrained $\Sigma_y = 1$. The validation results
are shown in Fig.\,\ref{fig_bw_valid}, and  in Table III that is shown in \cite{Tavares_2020_arxiv_comp}. Figure\,\ref{fig_bw_valid} shows a
free-noise simulation with the output prediction showed for two parts: (a) excited for $f_1 = 0.2$ with $0 \leq k \leq 16000$;
and (b) for $f_2 = 5$ with $ 16000\leq k \leq 18000$.

\subsection{Experimental Results}
\label{sec_exp_results}
\definecolor{cms}{HTML}{ff00ff}
\definecolor{cci}{HTML}{94a0a0}
\definecolor{inv}{HTML}{6cbbe7}

This section consider the identification process for a pneumatic valve. The NARX model obtained provides results that
are compared with models proposed by \cite{Rakotondrabe2011} and \cite{Abreu_etal2020}. See \cite{Tavares_2020_arxiv_comp}
for more details. The present pneumatic valve is the same used in \cite{Abreu_etal2020}, where the measured output is its
stem position and the input is a pressure signal applied to the valve after passing V/I and I/P conversion. The sampling
time is $T_{\rm s}=0.01\,{\rm s}$ and, for model identification, the same input designed in \cite{Abreu_etal2020} is used,
which is set as white noise low-pass filtered at $0.1\,{\rm Hz}$ similarly to what is done in the sections above, but for
only one frequency. For model validation, the inputs are sine waves with frequency $0.1\,{\rm Hz}$ and different amplitudes.
All these data sets are $200\,{\rm s}$ long ($N=20000$).

The first NARX model below is obtained for this system using the identification strategy based on the ERR
and AIC criterion, as explained in Sec.\,\ref{sp} and previous examples. The candidate
terms to compose these models are generated with $\ell = 3$, $n_y = 2$ and $n_u = 1$.

%
\vspace{0.2cm}
\noindent 1) ${\cal M}_{\rm h}$ describes the model identified with the inclusion of $\phi_{1}(k)$ and
$\phi_{2}(k)$, and also adopts the constraints proposed by \cite{Abreu_etal2020}. These constraints
are: the exclusion of some regressors showed in (i), (ii) and (iii), and force the parameters to comply with $\Sigma_y = 1$,
see Sec.\,\ref{sub_comp_hys}. The identified model is:
\begin{eqnarray}\label{mh_cns_valve}
y(k) & = & \hat{\rho}_1y(k-1)+\hat{\rho}_2y(k-2) + \hat{\rho}_3\phi_1(k-1) \nonumber \\
& + & {\rho}_4u(k-1)\phi_{1}(k-1)\phi_{2}(k-1) \nonumber \\ 
& + & \hat{\rho}_5y(k-2)\phi_{1}(k-1)\phi_2(k-1),
\end{eqnarray}
with $\hat{\rho}_1=9.76\times 10^{-1}$, $\hat{\rho}_2=2.40 \times 10^{-2}$, $\hat{\rho}_3=1.19 \times 10^{-1}$,
$\hat{\rho}_4=3.76$ and $\hat{\rho}_5=-4.73$. Note that, $\Sigma_y = \hat{\rho}_1 + \hat{\rho}_2 = 1$.

In what follows, three models already established in the literature are considered for comparison
purposes. The first one is a hysteresis model based on a phenomenological approach, which is stated
below.

\vspace{0.2cm}	
\noindent 2) ${\cal M}_{\rm bw}$ is used to represent a Bouc-Wen model previously shown in (\ref{eq_bw}).
To estimate the valve output, its parameters were re-estimated using an evolutionary approach based on
niches, which is formulated in \cite{tavares2019}. The obtained model is:
\begin{align}
\dot{h}(t) =& 7.54 \times 10^{-1}\dot{u}(t)-4.96|\dot{u}(t)|h(t)- 3.61 \dot{u}(t)|h(t)|,  \nonumber\\
y(t) =& 7.21\times 10^{-1}u(t) - h(t).
\label{eq_bw_val}
\end{align}

The last two models adopted were identified in \cite{Abreu_etal2020} for the same system under study and
with the same identification data. However, the meta-parameters of the identification process were $\ell = 3$, $n_y = 2$ and $n_u = 2$.

\vspace{0.2cm}
\noindent 3) ${\cal M}_{\rm h,2}$ symbolizes the model identified with the same constraints used for
${\cal M}_{\rm h}$ (\ref{mh_cns_valve}), plus an additional one required by the compensation strategy so
that the input signal can be isolated, see \cite{Abreu_etal2020}. The estimated model in the mentioned paper was:
\begin{align}
\label{Eq:Cms}
y(k)&=y(k-1)-19.76\phi_{1}(k-2) +19.32\phi_{1}(k-1) \nonumber\\
&{+}9.44\phi_{2}(k-2)\phi_{1}(k-2)u(k{-}2)\nonumber\\
&{-}12.61\phi_{2}({k{-}2})\phi_{1}(k{-}2)y(k{-}1).
\end{align}

\vspace{0.05cm}

\noindent 4) ${\breve{\cal M}}_{\rm h}$ represents the model identified to describe the inverse relationship
between $u(k)$ and $y(k)$ of the valve. Therefore, the model output is the estimated input $\hat{u}(k)$, the system
output $y(k)$ becomes the model input, and the candidate regressors $\breve{\phi}_{1}(k)=y(k)-y(k-1)$ and
$\breve{\phi}_{2}(k)={\rm sign}[\breve{\phi}_{1}(k)]$ are included. Following the same constraints used for
${\cal M}_{\rm h, cns}$ (\ref{mh_cns_valve}) with appropriate conversions for the inverse model, \cite{Abreu_etal2020} have got the following model:
\begin{align}
\label{Eq:Cci}
\hat{u}(k)&=\hat{u}({k{-}1})+86.67\breve{\phi}_{1}(k{-}1)-85.02\breve{\phi}_{1}(k{-}2) \nonumber \\ 
&{-}0.98\breve{\phi}_{1}(k{-}1)y({k{-}2}) +1.72\breve{\phi}_{2}(k{-}2)\breve{\phi}_{1}(k{-}2)y(k{-}2)  \nonumber \\ &{-}1.13\breve{\phi}_{2}(k{-}2)\breve{\phi}_{1}(k{-}2)\hat{u}(k{-}1).
\end{align}



 \begin{figure}[htb]
	\centering
	\includegraphics[width=1\columnwidth]{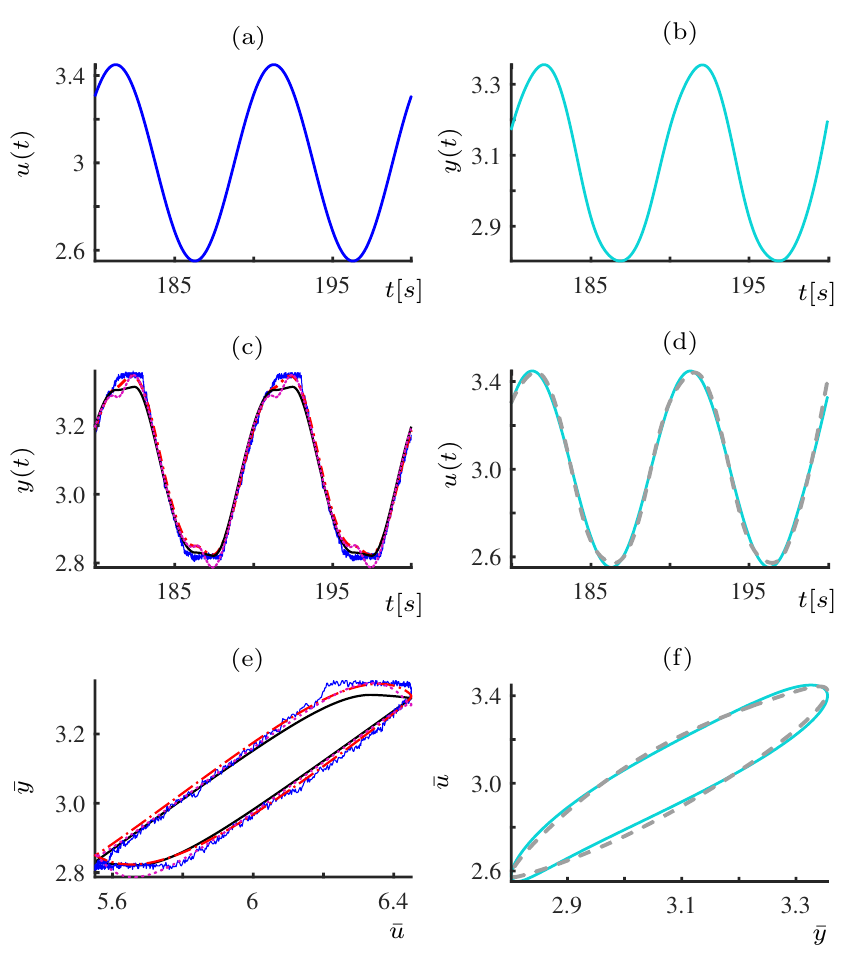}
	
	\vspace{-0.4cm}
	\caption{Free-run validation of the presented models obtained for the pneumatic valve. (a), (b) and (c) refer to the direct models which predict the output $y(t)$. In (c), the outputs $y(t)$ are shown for the models excited by the input (a), $u(t)=0.45{\rm sin}(2\pi (0.1)t+\pi/4)+3$V. (e) is the plane $\bar{u} \times \bar{y}$ which shows the hysteretic loop. (b), (d) and (f) are the analogous respectively for (a), (c) and (e), but to the inverse model  ${\breve{\cal M}}_{\rm h}$ (\ref{Eq:Cci}) (\textcolor{cci}{- -}) since it predicts the input $u(t)$ based on an smooth version of $y(t)$ showed in (b). (\textcolor{blue}{--}), refers to the system in the direct context while (\textcolor{inv}{--}), to the inverse. (\textcolor{red}{-$\vdot$-}), ${\cal M}_{\rm h}$ (\ref{mh_cns_valve}); (\textcolor{black}{--}), ${\cal M}_{\rm bw}$ (\ref{eq_bw_val}); (\textcolor{cms}{$\vdot$ $\vdot$}) ${\cal M}_{\rm h, 2}$, (\ref{Eq:Cms}).}
	\label{fig_exp_valid}
\end{figure}
 
The performance of each model is presented in Fig. \ref{fig_exp_valid}, considering a free-run simulation for $u(t)=0.55{\rm sin}(2\pi(0.1)t+\pi/4)+3$V that is showed in Fig.\,\ref{fig_exp_valid}-(a). In Fig.\,\ref{fig_exp_valid}-(b), the temporal evolution of the models' outputs are illustrated, while Fig.\,\ref{fig_exp_valid}-(c) presents the hysteretic curve in the plane $\bar{u} \times \bar{y}$. For a different perspective, the inverse model ${\breve{\cal M}}_{\rm h}$ is feed by a smooth version of the system output - Fig.\,\ref{fig_exp_valid}-(b) - to predict the corresponding input in the Fig.\,\ref{fig_exp_valid}-(d). The MAPE accuracy achieved for this figure is $1.8\%$ and the plane $\bar{y} \times \bar{u}$ is shown in Fig.\,\ref{fig_exp_valid}-(f). Other MAPE results are provided in Table V in \cite{Tavares_2020_arxiv_comp}.


\section{Conclusion}\label{Conclusion}

This work has presented three identified models for two numerical examples, and one experimental. An appropriate design of the excitation signal is provided since we select the frequencies that will be preserved and the operation regions to be achieved. The obtained models are simple, having no more than 5 terms. A natural step is use these models in the nonlinearity compensation context, which will be studied in \cite{Tavares_2020_arxiv_comp}.

\section*{Acknowledgment}

PEOGBA and LAA gratefully acknowledge financial support from CNPq
(Grant Nos. 142194/2017-4 and 303412/2019-4) and 
FAPEMIG (TEC-1217/98).	

\ifCLASSOPTIONcaptionsoff
  \newpage
\fi



\bibliographystyle{IEEEtran}
\bibliography{reference}
%

%
%
%
%
%

\end{document}